\documentclass[
   ,final            
  ]
  {aipproc}

\layoutstyle{6x9}

\begin{document}

\title{Low $Q^2$ Measurement of $g_2^p$ and the $\delta_{LT}$ Spin Polarizability}

                

\author{K. Slifer}{
  address={University of New Hampshire}
}

\begin{abstract}
JLab has been at the forefront of a program to measure
the nucleon spin-dependent structure functions 
over a wide kinematic range, and
data of unprecedented quality has been extracted in all three
experimental halls.
Moments of these quantities have proven to be powerful tools
to test QCD sum rules and  provide benchmark tests of
Lattice QCD and Chiral Perturbation Theory.
Precision measurements of $g_{1,2}^n$ and $g_1^p$ have been
performed as part of the highly successful `extended GDH program',
but data on the $g_2^p$ structure function remain scarce.
We discuss here JLab experiment E08-027, which will measure 
$g_2^p$
in the resonance region at low $Q^2$.
These data will be used to test the Burkhardt-Cottingham sum rule
and to extract the 
higher moments $\delta_{LT}^p(Q^2)$ and $\overline{d}_2^p(Q^2)$.
Data in the $Q^2$ range $0.02<Q^2<0.4$ GeV$^2$ will provide
unambiguous benchmark tests of $\chi$PT calculations on the lower end, while
probing  the transition region at the high $Q^2$ end where
parton-like behaviour begins to emerge.
This data will also have a significant impact on our theoretical understanding
of the hyperfine structure of the proton, and reduce the systematic uncertainty
of previous experiments which extracted the $g_1^p$ structure from purely longitudinal
measurements.
\end{abstract}

\maketitle

\section{Introduction}
Four independent structure functions are needed for a complete 
description of nucleon structure.  All spin-dependent effects are contained in $g_1$ and $g_2$, 
while spin-independent effects are parameterized in $F_1$ and $F_2$.
A simple physical interpretation is given in the impulse approximation of the parton model 
as distributions of quark momentum and spin in the nucleon:
\begin{eqnarray}
\nonumber
F_1(x) &=& \frac{1}{2} \sum e_i^2 \left[q_i(x)+ \bar{q}_i(x)\right]\\
F_2(x) &=& 2 x F_1(x) \\
\nonumber
g_1(x) &=& \frac{1}{2}  \sum e_i^2 \Delta q_i(x) 
\end{eqnarray}
But the parton model says nothing about the $g_2$ structure function.  
This fact, and the practical technical difficulty of producing transversely 
polarized targets has led to a historical neglect of the $g_2$ structure function.
In this proceedings we make the case that $g_2$ is a fundamental spin observable of the nucleon, and that  
measurements of $g_2$ give access to a wealth of interesting phenomenon.

\section{First Moments}
Comparison of the spin structure functions (SSF) to theoretical predictions is 
typically facilitated via the Cornwall-Norton (CN)~\cite{Cornwall:1968cx} moments:
\begin{eqnarray}
\nonumber
\Gamma_1^{(n)}(Q^2) = \int_0^1 dx~ x^{n-1} g_{1}(x,Q^2) \\
\label{eq:CN}
\Gamma_2^{(n)}(Q^2) = \int_0^1 dx~ x^{n-1} g_{2}(x,Q^2)
\end{eqnarray}
Here $x$ is the Bjorken scaling variable, and $Q^2$ represents the positive definite four momentum transfer of the virtual photon, which mediates the interaction between the incident electron probe and the nucleon target.  
By convention, the superscript is usually dropped in the case of $n=1$.
The first moment of $g_1$ represents the extended GDH integral, and has been investigated~\cite{Anthony:2000fn,Anthony:2002hy,Airapetian:2007mh,Amarian:2002ar,Amarian:2004yf,Meziani:2004ne,Osipenko:2004xg,Osipenko:2005nx,Slifer:2008re,Prok:2008ev,Solvigno:2008hk} with impressive precision over a wide kinematic range.


The first CN moment of the $g_2$ structure function is predicted to vanish by the Burkhardt-Cottingham (BC) sum rule~\cite{Burkhardt:1970ti}:
\begin{equation}
\label{eq:bc}
{\Gamma}_2 = \int_0^1 dx~g_2(x,Q^2) = 0
\end{equation}
This sum rule arises from the unsubtracted dispersion relation for the spin-dependent virtual-virtual Compton
scattering  amplitude $S_2$.  It is expected to 
be valid for any value of $Q^2$, although the reader is referred to~\cite{Jaffe:1989xx} for a detailed discussion of scenarios which would lead to violations of the relation.

\begin{figure}
   \includegraphics[height=.65\textheight]{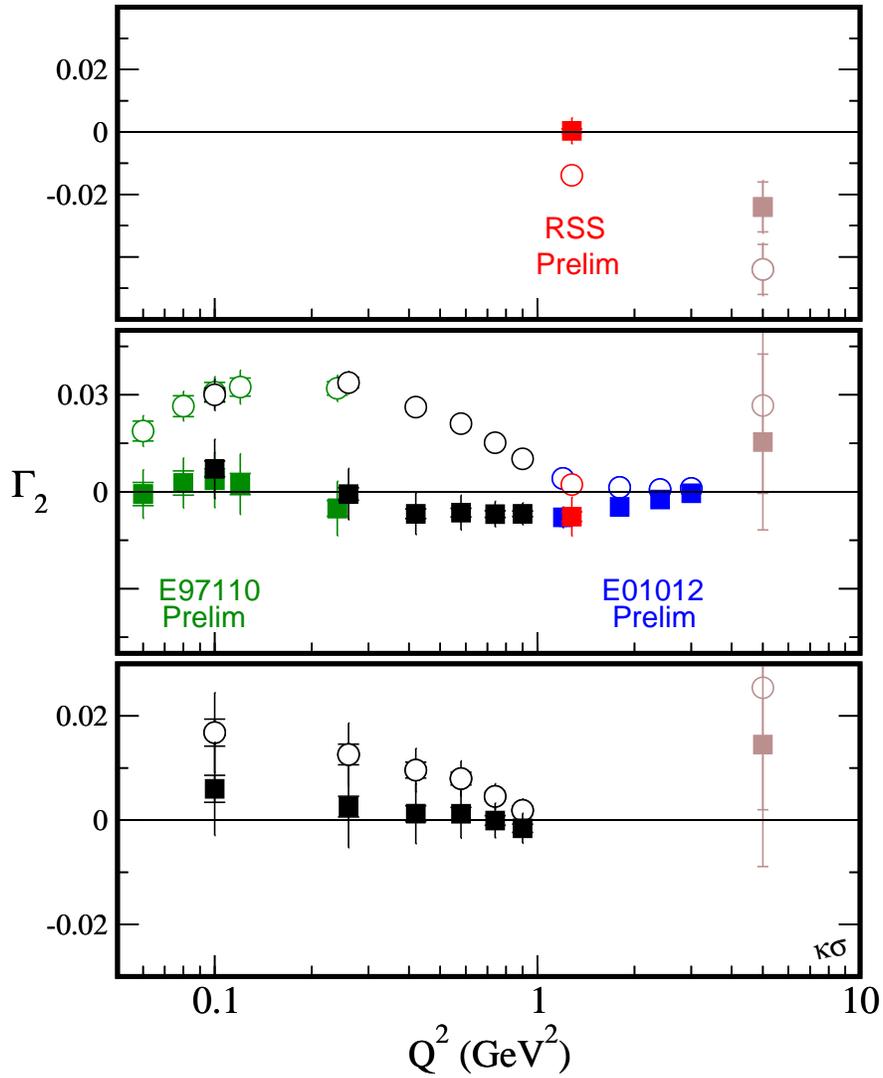}
   \caption{\label{BCWORLD}World data for $\Gamma_2(Q^2)$ for the proton (top), neutron (middle) and $^3$He (bottom). Open symbols represent measured data (typically the resonance region), while the full symbols include an estimate of the unmeasured contributions to the integral. Inner (outer) error bars represent statistical (total) uncertainties. 
Brown: E155 collaboration~\cite{Anthony:2002hy}. 
Red: RSS~\cite{Slifer:2008xu,Wesselmann:2006mw}. 
Black: E94010~\cite{Slifer:2008re,Amarian:2003jy}.
Green: E97110~\cite{SAGDH} (Very Preliminary). 
Blue: E01012~\cite{Solvigno:2008hk} (Very Preliminary).}
 \end{figure}

Fig.~\ref{BCWORLD} displays existing world data for $\Gamma_2(Q^2)$ for the proton, neutron and $^3$He.
All available data are consistent with the nuclear sum rule for $^3$He.  Similarly, we find satisfaction of the neutron sum rule, within uncertainties, across several different experiments and a large range of $Q^2$.  The picture is not so clear for the proton.
The E155 collaboration 
found their data to be inconsistent with the
proton BC sum rule at
$Q^2=5$ GeV$^2$, and the only other data point is from the RSS collaboration at $Q^2 = 1.3$ GeV$^2$.
%

The open symbols in Fig.~\ref{BCWORLD} represent the experimentally measured data which typically covers the resonance region.  To evaluate Eq.~\ref{eq:bc}, the contributions from $x=1$ and $x\to 0$ must be included.
The nucleon elastic contribution at $x=1$ can be easily evaluated using the form factor
parameterizations of~\cite{Lomon:2002jx,Mergell:1995bf}, and the nuclear elastic contribution for $^3$He is similarly determined from~\cite{Amroun:1994qj}.  All form factor parameterizations are known to high precision over the relevant region.

To estimate the $x\to 0$  contribution for the $g_2$ integrals,  
the Wandzura-Wilczek~\cite{Wandzura:1977qf} relation is used:
\begin{eqnarray}
\label{eq:WW}
g_2^{WW}(x,Q^2) \equiv -g_1(x,Q^2) + \int_x^1 \frac{g_1(y,Q^2)}{y} dy
\end{eqnarray}
This relation gives a prediction for the leading twist behaviour of $g_2$ entirely in terms of $g_1$.  
It is well known that existing data in the resonance region disagree with the Wandzura-Wilczek relation, but  
the higher twist effects which violate Eq.~\ref{eq:WW} are also expected to asymptotically vanish as 
$x\to 0$. 
Existing data~\cite{Kramer:2005qe,Slifer:2008xu} in the region of interest support this assumption.

\subsection*{Calculations of the Proton Hyperfine Structure}

\label{HYPERFINESECTION}
The first moments of the spin structure functions are also important input for calculations in 
atomic physics, and tests of physics beyond the standard model.
As discussed by Nazaryan, Carlson and Griffioen (NCG)~\cite{Nazaryan:2005zc,Carlson:2008ke},
the hyperfine splitting of hydrogen
has been measured
to a relative accuracy of 10$^{-13}$: 
\begin{eqnarray}
    \nonumber
    \Delta E = {\rm 1420.405\ 751\ 766\ 7(9)\ MHz} 
    \end{eqnarray}
but calculations of this fundamental quantity are only accurate to a
few parts per million.  This is due to the lack of knowledge of 
nucleon structure at low $Q^2$. 

\begin{figure}
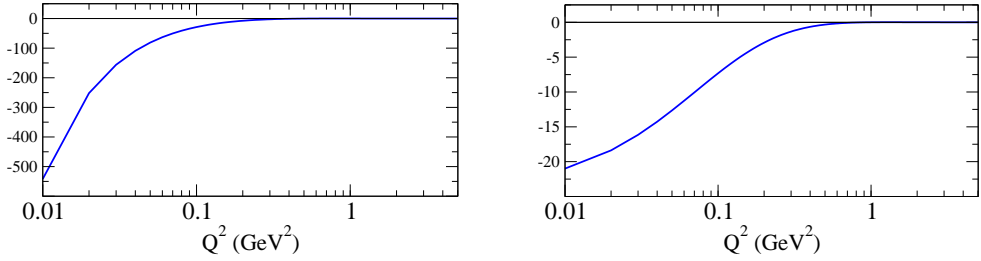

\includegraphics[width=0.410\textwidth]{figs/del1_small}
\hspace{0.8cm}
\includegraphics[width=0.4\textwidth]{figs/del2_small}
\caption{\label{HYPERPLOT}
MAID~\cite{Drechsel:1998hk} model prediction for the integrand
of $\Delta_1\simeq B_1/Q^4$ (left) and $\Delta_2\simeq B_2/Q^4$ (right). The scale for the 
horizontal axis is logarithmic.
}
\end{figure}

The splitting is conventionally expressed in terms of the
Fermi energy $E_F$ as
$\Delta E = (1+\delta) E_F$. 
where the correction $\delta$ is given by:
\begin{eqnarray}
\delta =
    1 + \left(\delta_{\rm QED}+\delta_R+\delta_{\varepsilon} \right) 
     +\Delta_S
    \end{eqnarray}
%
%
Here, $\Delta_S$ is the proton structure
correction and has the largest uncertainty. 
Recoil effects are accounted for in
$\delta_R$,
and 
$\delta_{\rm QED}$ represents the QED radiative correction which
is known to very high accuracy. 
The relatively small corrections for hadronic and muonic vacuum polarizations,  along with
the weak interaction 
are collected 
into $\delta_{\varepsilon}$. 
%
The structure-dependent term $\Delta_S$ 
depends on ground state and excited properties of the proton.
It is conventionally split
into two terms:
$\Delta_S = \Delta_Z + \Delta_{\rm pol}$, 
%
where the first term can be determined by measurements of the Zemach radius~\cite{Zemach} 
in elastic scattering. 
%
%
%
%
%
%
%
%
The second term, $\Delta_{\rm pol}$, 
involves contributions
where the proton is 
excited. 
\begin{equation}
\label{DELL2}
\Delta_{\rm pol} \sim
    (\Delta_1+\Delta_2)
    \end{equation}
$\Delta_1$ involves the inelastic Pauli form factor $F_2$ and the $g_1$ structure function,
while  $\Delta_2$ depends only on the $g_2$ structure function:
\begin{eqnarray}
  \Delta_1 &=& \frac{9}{4}\int_0^\infty \frac{dQ^2}{Q^2}\left\{F_2^2(Q^2) + \frac{8m_p^2}{Q^2}
B_1(Q^2)\right\}   
										\nonumber\\
\Delta_2 &=& -24m_p^2\int_0^\infty \frac{dQ^2}{Q^4}B_2(Q^2) 
	\label{eq:Delta}
    \end{eqnarray}
The integrals $B_1$ and $B_2$ are very similar to the first moments of Eq.~\ref{eq:CN},
\begin{eqnarray}
B_1(Q^2) &=& \int_0^{x_{\rm th}} dx \, \beta_1(\tau) g_1(x,Q^2)    \nonumber \\
B_2(Q^2) &=& \int_0^{x_{\rm th}} dx \, \beta_2(\tau) g_2(x,Q^2)  
\label{eq:B}
\end{eqnarray}
but involve some additional kinematic factors $\beta_1$ and $\beta_2$. 
\begin{eqnarray}
\beta_1(\tau) 
&=& 
=
\frac{4}{9} \left[
        - 3\tau + 2\tau^2 + 2(2-\tau)\sqrt{\tau(\tau+1)}  \right] 
                \nonumber \\
\beta_2(\tau) 
&=& 
1+2\tau-2\sqrt{\tau(\tau+1)}
    \end{eqnarray}
Here $\tau = {\nu^2} \big/ {Q^2}$, and
$x_{th}$ represents the pion production threshold. 
The $Q^2$ weighting of $\Delta_1$ and $\Delta_2$ ensures that the
low momentum transfer region dominates these integrals as demonstrated in Fig.~\ref{HYPERPLOT}.
Because of this kinematic weighting, precise measurements of $g_1$ and $g_2$ at low $Q^2$ 
can have significant impact on calculations of the hydrogen hyperfine splitting.

\subsection{Higher Moments}
Moments of the spin structure functions that involve
higher powers of the Bjorken variable $x$ are also of great interest.  
These moments serve as powerful tools for
theoretical analysis, and have one very practical advantage over the first moments
of the previous section:
the additional $x$-weighting reduces uncertainties associated with the low-x extrapolations.
One example is the forward spin polarizabilities of the nucleon. 
The electric and magnetic polarizabilities measure
the nucleon's response to an external electromagnetic
field. 
The
extension of these quantities to 
finite 
$Q^2$ leads to the concept of the generalized polarizabilities~\cite{Guichon:1995pu},
which can be related to the scattering amplitudes for
 forward
virtual Compton scattering (VCS) and forward doubly-virtual Compton scattering (VVCS).
With this additional dependence on $Q^2$,
the generalized polarizabilities provide a powerful tool
to probe the nucleon structure covering the whole range
from the partonic to the hadronic region. In particular,
the generalized polarizabilities provide one of the most
extensive tests of $\chi$PT calculations in the low $Q^2$ 
region.

Of particular interest are the forward polarizabilities $\gamma_0$ and $\delta_{LT}$, which 
can be evaluated from  the spin structure
functions $g_1$ and $g_2$: 
\begin{eqnarray}
\gamma_0(Q^2)&=&
%
\label{GAM0DEF}
\frac{16 \alpha M^2}{Q^6}\int^{x_0}_0 x^2 \Bigl [g_1(x,Q^2)-\frac{4M^2}{Q^2}
x^2g_2(x,Q^2)\Bigr ] dx. \\
%
\delta_{LT}(Q^2)&=&
%
\label{DLTDEF}
\frac{16 \alpha M^2}{Q^6}\int^{x_0}_0 x^2 \Bigl [g_1(x,Q^2)+g_2(x,Q^2)
\Bigr ] dx.
\end{eqnarray}
and which, in principle, should be ideal quantities to test calculations of
chiral perturbation theory ($\chi$PT) at low $Q^2$.
%

%
%
%
%
%
%



One additional higher moment  has played an important role in recent investigations:
\begin{eqnarray}
I(Q^2) = \int_0^1 x^2\left[2 g_1(x,Q^2) + 3 g_2(x,Q^2)\right] dx
\end{eqnarray}
This moment provides one method to access what are known as `higher twist' effects, which
arise from multi-parton interactions and the finite (non-zero) value of quark masses. These effects manifest as
deviations from the predictions of the parton model. 
At very large $Q^2$, $I(Q^2)$ can be related to the twist-3 matrix element $d_2$ in  the Operator Product 
Expansion (OPE).  
The physical interpretation of this quantity is quite interesting:
$d_2$ has been related~\cite{Filippone:2001ux} to the
color polarizabilities, 
which describe how the color electric and magnetic fields
respond to the nucleon spin.  More recently, an alternate description has emerged~\cite{Burkardt:2009rf} which identifies $d_2$ with the transverse component of the color-Lorentz force
acting on the stuck quark in the instant after absorbing the virtual
photon.

The definition of I($Q^2$) as a higher moment 
holds for all $Q^2$.
Considering that it must vanish in the limits $Q^2\to 0$ and $Q^2\to\infty$, 
but peaks around 1 GeV$^2$, it serves as a measure of QCD complexity,
and provides a unique tool to study
the transition from perturbative to non-perturbative behaviour.
See Refs.~\cite{Osipenko:2005nx,GRIFF} for an interesting discussion of I($Q^2$) at low and
moderate $Q^2$.

\subsection{The need for more $g_2^p$ data}

\begin{figure}
\includegraphics[width=0.5\textwidth]{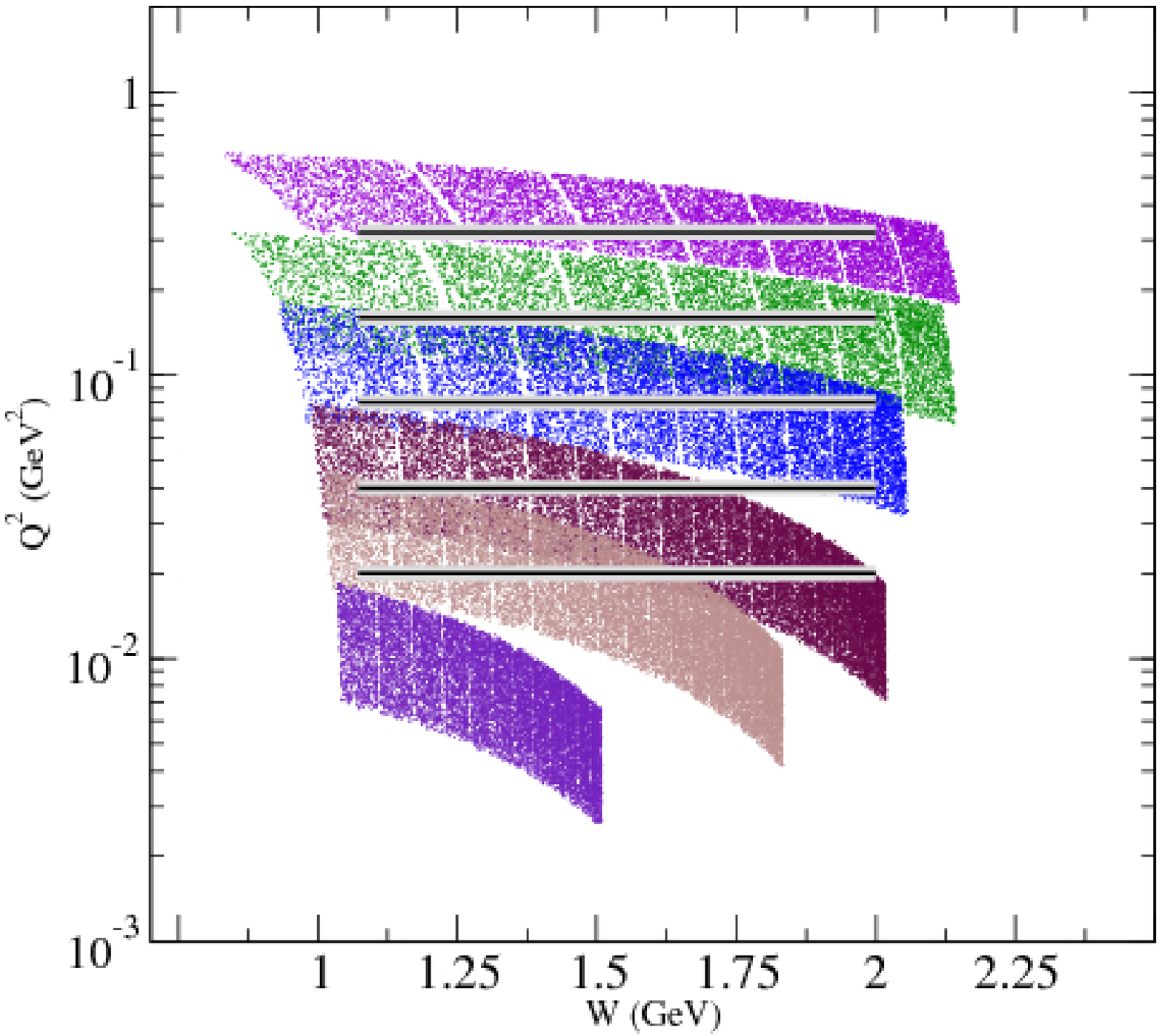}
\includegraphics[width=0.5\textwidth]{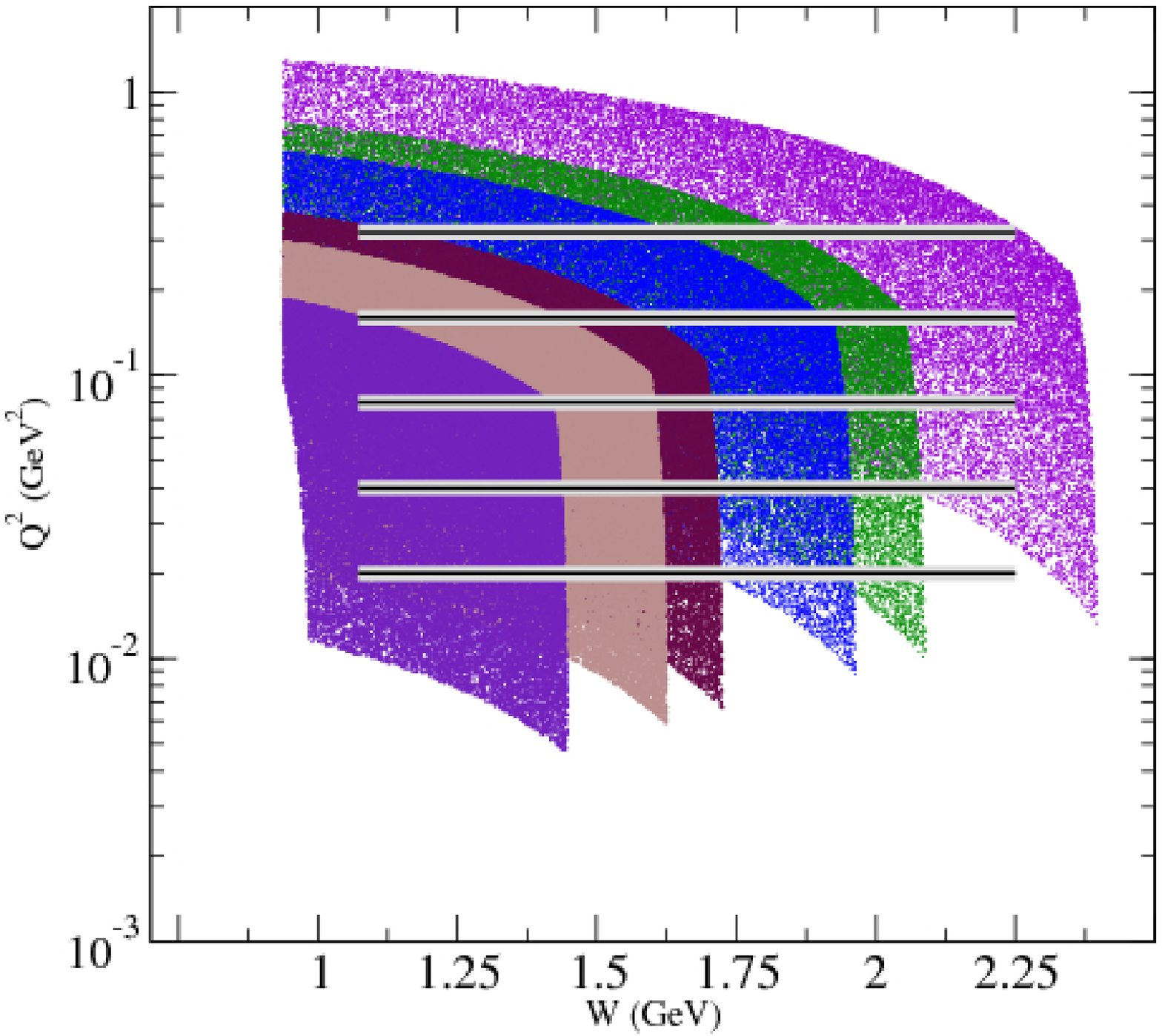}
\caption{
Kinematical coverage in $Q^2$ (GeV$^2$) vs. W (GeV). 
{\bf Left:} Hall A E08-027. Measurement of $g_2^p$ and $\delta_{LT}$.
Each band represents a planned incident energy and scattering angle combination.
$E_0$ = 4.4, 3.3, 2.2, 1.1 GeV, with $\theta=6$ and $9$ degrees.
Horizontal lines represent values of constant $Q^2$.
{\bf Right:}  Hall B EG4.   Measurement of $g_1^p$ and $g_1^d$.  Incident energy $E_0=$ 3.0, 2.3, 2.0, 1.5, 1.3, and 1.0 GeV. 
Proton target was utilized at all energies.  Deuteron target
utilized for $E_0=1.3$ and $2.0$ GeV.
}
\label{DLTKIN}
\end{figure}

\label{DLT}
At low and moderate $Q^2$,
data on the
$g_2^p$ structure function is conspicuously absent.
The lowest momentum transfer that has been investigated is
$1.3$ GeV$^2$ by the RSS collaboration~\cite{Wesselmann:2006mw}. 
The absence of $g_2^p$
data is particularly unsatisfying given the intriguing results found in the
transverse neutron data: The E155
collaboration~\cite{Abe:1998wq}
found their data to be inconsistent with the proton Burkhardt-Cottingham (BC) sum rule~\cite{Burkhardt:1970ti} at
$Q^2=5.0$ GeV$^2$, while the E94-010
collaboration~\cite{Amarian:2004tmp}
found that the
neutron BC sum rule held below
$Q^2=1.0$ GeV$^2$.
Even more compelling, it was found that state-of-the-art 
NLO
$\chi$PT calculations are in agreement
with  data for the generalized polarizability $\gamma_0^n$ at $Q^2=0.1$ GeV$^2$, but
exhibit a significant discrepancy~\cite{Amarian:2004yf} with the longitudinal-transverse polarizability
$\delta_{LT}^n$.
This is  surprising since
$\delta_{LT}$ 
is insensitive to the dominating $\Delta$ resonance contribution.
For this reason, it was believed that $\delta_{LT}$ should 
be more suitable than
$\gamma_0$ to serve as a testing ground for the chiral dynamics of
QCD~\cite{Bernard:2002bs,Kao:2002cp}. 
It is natural to ask if this discrepancy exists also in the proton case,
and determining
the isospin
dependence will help to solve this puzzle.
In addition, $\chi$PT  
is now being used to help Lattice QCD
extrapolate to the physical region,
so it is critical to have benchmark tests of the reliability of existing 
calculations.

Also, as discussed above, 
lack of knowledge of the $g_2^p$ structure function at low $Q^2$
is one of the leading uncertainties in ongoing calculations
of the hyperfine splitting of the hydrogen atom. 
In particular, Fig.~\ref{HYPERPLOT} reveals that $\Delta_2$ is dominated by the contribution
below $0.4$ GeV$^2$ where no data exists. 

\begin{figure}
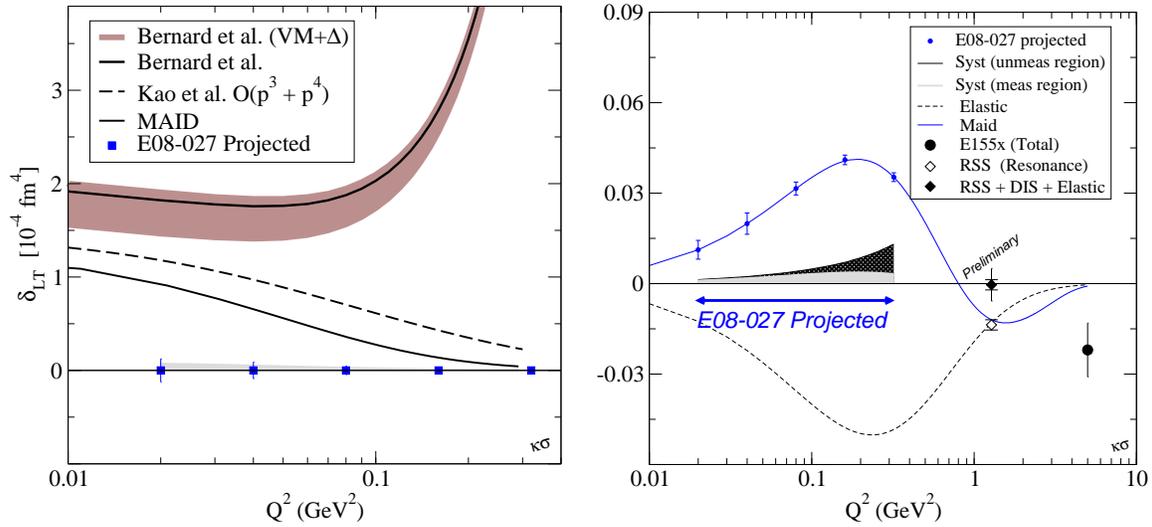

\includegraphics[width=0.5\textwidth]{figs/dlt_bernard_log}
\hspace{0.2cm}
\includegraphics[width=0.5\textwidth]{figs/gam2_log_07001}
\caption{\label{PRED1}
{\bf Left:}  Projected uncertainties for $\delta_{LT}$.
$\chi$PT predictions from Bernard {\it et al.}~\cite{Bernard:2002bs},
and Kao {\it et al.}~\cite{Kao:2002cp}.
{\bf Right:}
Projected uncertainties for $\Gamma_2^p(Q^2)$.
The light and dark bands on the horizontal axis represent the
experimental systematic, and the uncertainty arising from the unmeasured
($x \to 0$, $x=1$) contributions to $\Gamma_2$, respectively.}
\end{figure}

\begin{figure}
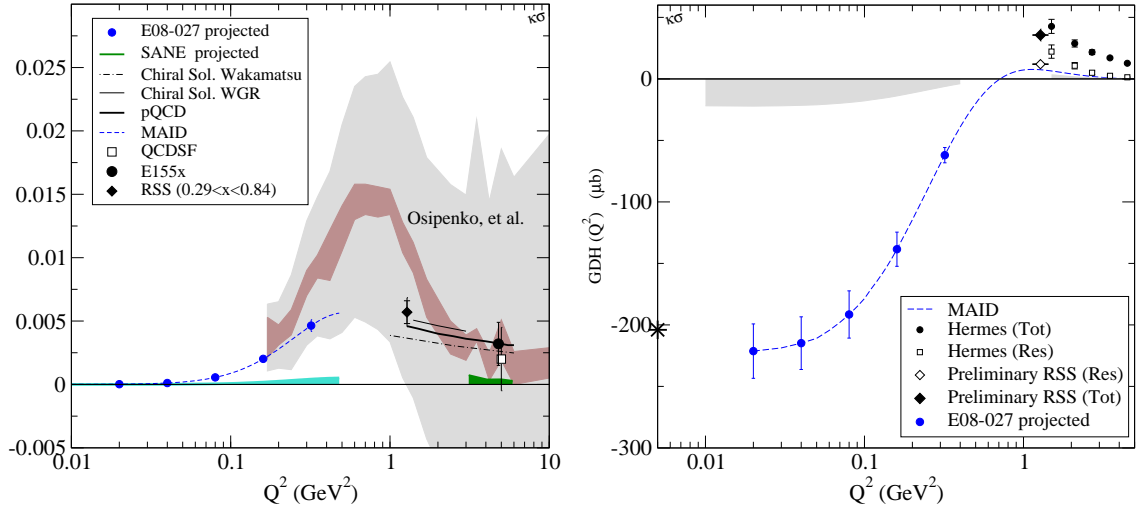

\includegraphics[width=0.5\textwidth]{figs/d2_expected_new}
\hspace{0.05cm}
\includegraphics[width=0.5\textwidth]{figs/gdhp_b_scaled_new}
\caption{\label{PRED2}
{\bf Left:} Projected uncertainties for I($Q^2$). 
    Statistical errors are shown on the symbols.
    Systematic is represented by the band on the axis.
    MAID~\cite{Drechsel:1998hk} model.
    Expected SANE uncertainties shown on the horizontal axis.
    PQCD from Ref.~\cite{Shuryak:1981pi}.
    Lattice QCD calculation from~\cite{Gockeler:2000ja}.
    WGR chiral soliton model from Ref.~\cite{Weigel:2000gx}.
    Wakamatsu's chiral soliton model from Ref.~\cite{Wakamatsu:2000ex}.
    The inner (outer) band of the
    Osipenko {\it et al.} global analysis~\cite{Osipenko:2005nx}:
    represents statistical (systematic) uncertainty.
    Unmeasured non-resonance contribution is highly suppressed by $x^2$ weighting of $I(Q^2)$
    and is not shown.
{\bf Right:}     Projected results for the GDH Integral $GDH(Q^2)$. Statistical errors are shown on the symbols.  Systematic is represented by the band
on the axis.
Also shown is HERMES~\cite{Airapetian:2002wd}, and RSS preliminary data.
}
\end{figure}

\subsection{E08-027: The $g_2$ Structure Function and the LT Spin Polarizability}
E08-027 was approved to run in JLab Hall A with A$^-$ rating for 24 days.
This experiment will  perform an inclusive  measurement at forward angle of the proton
spin-dependent cross sections in order to determine the
$g_2^p$ structure function in the resonance region for $0.02<Q^2<0.4$ GeV$^2$.
From these data we can evaluate the BC sum, and the longitudinal-transverse polarizability $\delta_{LT}$.
The kinematic coverage is shown in Fig.~\ref{DLTKIN}, where it is compared to Hall B experiment EG4, which performed a measurement of $g_1^p$ and $g_1^d$ in a similar kinematic region.

E08-027 will require a major installation that
involves significant changes to the existing Hall A beamline in order
to properly transport the electron beam in the presence of the 5T
magnetic field of the transversely polarized target.
For this experiment we will install the JLab-UVA polarized
ammonia target in Hall A for the first time.
This target exploits the
Dynamical Nuclear Polarization (DNP) technique to polarize a solid ammonia insert
maintained in a liquid helium bath at 1 K in a 5 Tesla
field.
This installation has the potential to initiate  a very broad  spin
structure program in Hall A that would extend naturally into the 12 GeV era.

Projected results for the longitudinal-transverse polarizability $\delta_{LT}$ and the BC sum rule
are shown in Fig.~\ref{PRED1}. Fig.~\ref{PRED2} shows the precision expected on the $I(Q^2)$ integral
and the extended GDH sum rule.
To assess the impact of E08-027 on calculations of the hyperfine splitting, we consider the existing calculations.  NCG~\cite{Nazaryan:2005zc} utilized a model for $g_2^p$ to obtain:
$\Delta_{pol} = (1.3\pm 0.3)~\textrm{ppm}$,
of which 0.13 ppm uncertainty arises from an assumed 100\% uncertainty on $g_2^p$. 
The total uncertainty  projected for
E08-027 is better than 10\%, so {naively} we might expect
the published error on $\Delta_2$ to improve by an order of magnitude from
$\pm 0.57$ to $\pm 0.06$, and the error contribution of $g_2$ to $\Delta_{pol}$
to decrease by an order of magnitude from 0.13 ppm to 0.013 ppm.  
However, 
comparison to existing $g_2$ data reveals that
none of the available models 
are strongly favored.  
We find that the variation among model predictions for $\Delta_2$ is much larger than 100\%. MAID\footnote{Integrated over the region $W\le 2$ GeV and $Q^2\le 5$
GeV$^2$. }
predicts $\Delta_2=-1.98$,
while
the CLAS model 
and the Simula model predict $\Delta_2=-0.57\pm 0.57$, and
$-1.86\pm 0.37$ respectively.
The disparity among existing model predictions is quite large,
which is natural considering the lack of data in this region.  
As such, E08-027 will provide the first realistic determination of $\Delta_2$.



\section{Conclusion}
Accurate measurement of the spin observables $g_1$ and $g_2$ is necessary to
obtain a quantitative understanding of the nucleon.
We described here an upcoming experiment that will provide much needed information on the
relatively unknown spin structure function $g_2^p$.
These data will
significantly impact ongoing calculations of the hydrogen hyperfine splitting,
and provide benchmark tests of state-of-art chiral perturbation theory calculations.
In addition, they will 
allow  tests of the extended Gerasimov-Drell-Hearn, and the Burkhardt--Cottingham sum rules.

\begin{theacknowledgments}
I would like to thank the spokesmen of RSS: Mark Jones and Oscar Rondon,  E97110: 
Alexandre Deur, J.P. Chen and Franco Garibaldi,  and E01012: Nilanga Liyanage, J.P Chen and Seonho Choi for kindly allowing me to present their preliminary data here.  Many thanks also to Vince Sulkosky and Patricia Solvignon for providing the raw data.
This work was supported by Department of Energy contract DE-FG02-88ER40410.
\end{theacknowledgments}

\bibliographystyle{aipproc}   

\bibliography{slifer_ssld}

\end{document}